\documentclass[prl,aps,epsfig,twocolumn]{revtex4}

\usepackage{epsfig}

\def\ba{\begin{eqnarray}}
\def\ea{\end{eqnarray}}
\def\beq{\begin{equation}}
\def\eeq{\end{equation}}

\newlength{\textwidthm}
\begin{document}
\title{Stability of atomic clocks based on entangled atoms}
\author{A. Andr\'{e}, A. S. S\o rensen, and M. D. Lukin}
\affiliation{Physics Department and Institute for Theoretical Atomic and 
Molecular Physics, \\ Harvard University, Cambridge, Massachusetts 02138}
\date{\today}

\begin{abstract}
We analyze the effect of realistic noise sources for an atomic clock 
consisting of a local oscillator that is actively locked to 
a spin-squeezed (entangled) ensemble of $N$ atoms. 
We show that the use of entangled states can lead to an improvement
of the long-term stability of the clock when the measurement is limited by 
decoherence associated with instability of the local oscillator combined 
with fluctuations in the atomic ensemble's Bloch vector.
Atomic states with a moderate degree of entanglement yield the maximal 
clock stability, resulting in an improvement that scales as $N^{1/6}$ 
compared to the atomic shot noise level. 
\end{abstract}
\maketitle

Quantum entanglement is the basis for
many of the proposed applications of quantum information science
\cite{bouwmeester}. 
The experimental implementation of these ideas is challenging since
entangled states are easily destroyed by decoherence.
To evaluate the potential usefulness of entanglement it is therefore 
essential to include a realistic description of noise in experiments of 
interest. 
Although decoherence is commonly analyzed in the context of 
simple models \cite{gardiner},
practical sources of noise often possess a non-trivial frequency
spectrum, and enter through a variety of different physical processes. 
In this Letter, we analyze the effect of realistic decoherence 
processes and noise sources in an atomic clock that is actively locked to 
a spin-squeezed (entangled) ensemble of atoms.

The performance of an atomic clock can be
characterized by its frequency accuracy and
stability. Accuracy refers to the frequency offset from the ideal
value, whereas stability describes the fluctuations around, and drift away 
from the average frequency. 
To improve the long-term clock stability, it has been 
suggested to use entangled atomic ensembles  
\cite{wineland93,wineland94,wineland96}, and in 
this letter we analyze such proposals in the presence of 
realistic decoherence and noise. 
In practice, an atomic clock operates by
locking the frequency of a local     
oscillator (L.O.) to the transition frequency between two
levels in an atom. This locking is achieved by a spectroscopic measurement
determining the 
L.O. frequency offset $\delta \omega$ from the atomic resonance, followed
by a feedback mechanism which steers the L.O. frequency so as to null the 
mean frequency offset.
The problem of frequency control thus combines elements of 
quantum parameter estimation theory and control of stochastic systems via 
feedback \cite{mabuchi,wiseman}.

The spectroscopic measurement of 
the atomic transition frequency is typically achieved through Ramsey
spectroscopy \cite{ramsey}, in which the atoms are illuminated by two 
short, near-resonant pulses from the local 
oscillator, separated by a long period of free
evolution, referred to as the Ramsey time $T$.
During the free evolution the atomic state and the L.O.
acquire a relative phase difference $\delta\phi=\delta\omega T$, which is
subsequently determined by a projection measurement. If 
a long time $T$ is used, then Ramsey spectroscopy
provides a very sensitive measurement of the L.O. frequency offset 
$\delta\omega$ \cite{footnote1}. 
Here, we investigate the situation relevant to trapped 
particles, such as atoms in an optical lattice \cite{katori} or trapped 
ions \cite{wineland00}. 
In this situation, the optimal value of $T$
is determined by atomic decoherence (caused by 
imperfections in the experimental setup) which therefore determines the 
ultimate performance of the clock.

We consider an ensemble of $N$ two-level particles with
lower (upper) state $|\downarrow\rangle$ ($|\uparrow\rangle$). 
Adopting the nomenclature of spin-1/2 particles, we introduce the total 
angular momentum (i.e., Bloch vector)
$\vec{J}=\sum_{j=1}^{N}\vec{S}_j$, where e.g.
$S_z^j=\left(|\uparrow\rangle_j\langle\uparrow|-
|\downarrow\rangle_j\langle\downarrow|\right)/2$. Initially the state of
the atoms has mean $\langle\vec{J}\rangle$ along the $z$ direction 
and $\langle J_x\rangle=\langle J_y\rangle=0$.
Unavoidable fluctuations in the
$x$ and $y$ components $\langle J_x^2\rangle=\langle J_y^2\rangle\neq 0$,
result in the so-called atomic projection noise.
These fluctuations give rise to an uncertainty in the Ramsey phase
$\delta\phi_R\simeq\Delta J_y/|\langle\hat{J}_z\rangle|$ as indicated
geometrically in Fig. 1 \cite{santarelli99,wineland93}.  
For uncorrelated spins aligned along the $z$ axis, the
uncertainty from independent spins are added in quadrature, resulting in
the projection noise $\Delta J_y=\sqrt{N}/2$. To reduce
the measurement error it has been proposed 
\cite{wineland94,kitagawa93,sorensen01} and demonstrated \cite{meyer01}
to use entangled atomic states (so called spin squeezed states), which 
have reduced noise in one of the transverse spin components (e.g., 
$J_y$) and non-zero noise $\Delta J_z$ in the mean spin direction.
Ideally this gives an improvement by a factor $\xi=\sqrt{N}\Delta 
J_y/\langle J_z\rangle$, which
can be as low as $\xi=1/\sqrt{N}$ for maximally entangled states
\cite{wineland96}.

Using a simple noise model it was shown in Ref.\ \cite{huelga97} that 
entanglement provides little gain in spectroscopic sensitivity in the 
presence 
of atomic decoherence. In essence, random fluctuations in the phase of the 
atomic coherence cause a 
rapid smearing of the error contour in Fig. 1a.
For example, dephasing of individual particles results in an additional 
contribution $(N/4)\langle\delta\phi^2\rangle$ to the noise, 
where $\langle\delta\phi^2\rangle$ denotes the variance of the phase 
fluctuations 
(increasing with $T$ as $\langle\delta\phi^2\rangle=\gamma T$ for 
white noise, where $\gamma$ is the dephasing rate).
In practice, the stability of atomic clocks is often limited primarily by
fluctuations of the L.O.
As we show below, 
the L.O. fluctuations result in the added noise
$\Delta J_z^2\langle\delta\phi^2\rangle$,
where $\Delta J_z^2$ is the initial variance in $J_z$.
This added noise is due to the error in the feedback loop, caused by 
the longitudinal noise $\Delta J_z$.
For weakly entangled states, the added noise is considerably smaller 
than in the case of atomic dephasing
and the use of entangled states
can lead to a significant improvement in clock stability. 

In what follows we outline a model that incorporates the effects of atomic 
noise and spin squeezing as well as that of the feedback loop. 
Before proceeding, we note that  qualitative considerations  along these lines
were noted in Ref.\ \cite{winelandbible}.
At the operating point, the error signal in Ramsey 
spectroscopy \cite{wineland94} measured at time $t_k$ is determined by the 
operator
\beq
\hat{E}(t_k)\simeq\sum_{j=1}^N
\hat{S}_z^j\sin[\delta\phi_j(t_k)]+\hat{S}_y^j\cos[\delta\phi_j(t_k)],  
 \label{eq:nonlinear}
\eeq
where $\delta\phi_j(t_k)$
is the phase acquired by the $j$th atom 
during the interrogation time $T$ and all operators refer to the initial 
atomic state.
We separate the phase
into two parts $\delta\phi_j=\delta\phi_O+\delta\phi_E^j$, where
$\delta\phi_O(t_k)=\int_{0}^{T}\delta\omega(t_k-t)dt$ is the phase due to 
the frequency fluctuations $\delta \omega(t)$ 
of the L.O., and
$\delta\phi_E^j$ is a phase induced by
the interaction of the $j$th atom with the environment. 
In order to lock the L.O. to the
atomic frequency, the interrogation time should be short enough that
$\langle\delta\phi(t_k)^2\rangle\lesssim 1$.
Expanding in terms of $\delta\phi(t_k)$, we find the measured error signal 
\ba
\mathcal{E}(t_k) &\simeq& \langle\hat{J}_z\rangle
\left(\delta\phi_O(t_k)-\frac{\delta\phi_O(t_k)^3}{3!}\right)
+\sum_{j=1}^{N}\mathcal{S}_z^j\delta\phi_E^j(t_k)
\nonumber \\
&+& 
[\mathcal{J}_y(t_k)+\delta\mathcal{J}_z(t_k)\delta\phi_O(t_k)]+\cdots.
\label{eq:linear}
\ea
Here $\delta\mathcal{J}_z(t_k)=\mathcal{J}_z(t_k)-\langle\hat{J}_z\rangle$,
where $\mathcal{J}_z(t_k)$ and $\mathcal{J}_y(t_k)$
are random numbers with a 
distribution corresponding to the initial atomic state (we 
consider here states for which $\langle\hat{J}_y\hat{J}_z\rangle=0$, so 
that we may treat $\mathcal{J}_z(t_k)$ and $\mathcal{J}_y(t_k)$ as 
independent random variables).
The term multiplying $\langle\hat{J}_z\rangle$ in 
(\ref{eq:linear}) is used to estimate the frequency 
offset, while the remaining terms represent measurement noise. 

The feedback is started at $t=0$ and, at the end of each Ramsey cycle, at 
$t_k=kT$ ($k=1,2,...$), the detection
signal is used to steer the frequency of the oscillator to
correct for the fluctuations accumulated during the last cycle
$\delta\omega(t_{k}^+)=\delta\omega(t_k^-)+\Delta\omega(t_k)$, where
$t_k^-$ and $t_k^+$ refer to before and after the correction, and 
$\Delta\omega(t_k)$ is the frequency correction. Assuming that negligible 
time is spent performing the $\pi/2$ pulses and in preparing and detecting 
the state of the atoms, the mean frequency offset after
running for a period  $\tau=nT$ is then
\ba
\delta{\bar \omega}(\tau) &=& 
\frac{1}{\tau}\int_{0}^{\tau}\delta\omega(t)dt 
\nonumber \\
&=& 
\frac{T}{\tau}
\sum_{k=1}^{n}\left[\frac{\delta\phi_O(t_k)}{T}+\Delta\omega(t_k)\right].
\label{eq:meanfreq}
\ea

We begin by analyzing the simplest case of linear feedback (in 
$\mathcal{E}(t_k)$)
and later extend to the more optimal nonlinear feedback case.
With $\Delta\omega(t_k)=\frac{-\mathcal{E}(t_k)}{\langle\hat{J}_z\rangle 
T}$, using (\ref{eq:linear}) and substituting
in (\ref{eq:meanfreq}), we find, ignoring for now the 
$\delta\phi_O(t_k)^3$ term,
\ba
\delta{\bar \omega}(\tau) &=&
\frac{-1}{\tau\langle\hat{J}_z\rangle}\sum_{k=1}^{n}
[\mathcal{J}_y(t_k)+\delta\mathcal{J}_z(t_k)\delta\phi_O(t_k) 
\nonumber \\
&+& 
\sum_{j=1}^{N}\mathcal{S}_z^j\delta\phi^j_E(t_k)]
\label{eq3}
\ea
Note that the acquired
offsets $\delta\phi_O(t_k)/T$ ($k=1,...,n$) due to L.O. frequency 
fluctuations are corrected by the 
feedback loop and do not appear in (\ref{eq3}),
while measurement noise is added at the detection times $t_k$. 
The first two terms in (\ref{eq3}) are uncorrelated for
different $t_k$ since the atomic noise for different detection events is
uncorrelated. 
If the dephasing noise is uncorrelated for different $t_k$,
then the fractional frequency fluctuation (Allan deviation) \cite{barnes71}
$\sigma_y(\tau)=\langle(\delta{\bar \omega}(\tau)/\omega)^2\rangle^{1/2}$,
is 
\beq
\sigma_y(\tau) =
\frac{
\left[\Delta J_y^2+\Delta J_z^2\langle\delta\phi_O^2\rangle+
(\lambda \langle J_z ^2\rangle)\langle\delta\phi_E^2\rangle\right]^{1/2}}
{\omega\sqrt{\tau T}\langle\hat{J}_z\rangle}.
\label{eq:sigmay}
\eeq
Here $\lambda$ accounts for the possibility of collective
decoherence, so that for atoms
dephasing collectively (independently) $\lambda\rightarrow 1$
($\lambda\rightarrow(N/4)/\langle\hat{J}_z^2\rangle$).
The L.O. noise affects the atoms in a similar fashion than collective 
dephasing.
Note, however, the significant difference between 
collective environmental dephasing, 
which enters expression (\ref{eq:sigmay}) as
$\langle\hat{J}_z^2\rangle\langle\delta\phi_E^2\rangle$, and 
L.O. noise, for which $\Delta J_z^2\langle\delta\phi_O^2\rangle$ is the
relevant expression. 
The feedback loop results in a large cancellation of
the effect of the L.O. noise on the stability;
the uncanceled part of the noise is
now proportional to $\Delta J_z^2\ll\langle\hat{J}_z^2\rangle$.
\begin{figure}
\epsfig{file=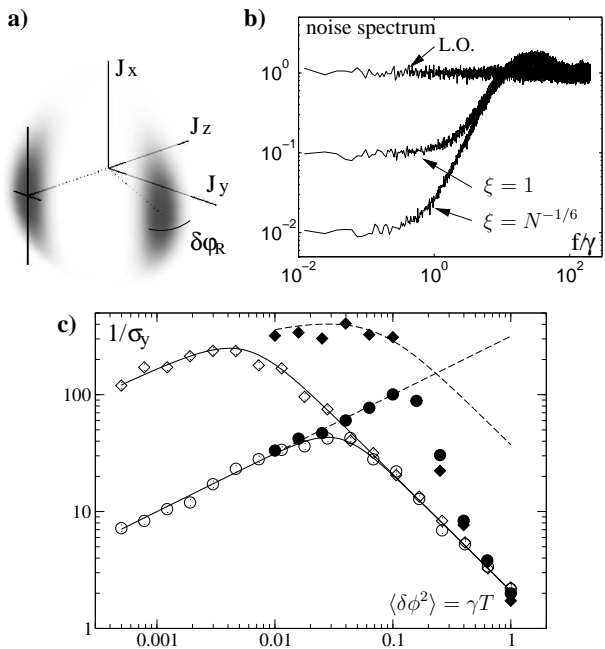,width=8.0cm}
\caption{
a) Representation of the probability distribution on the Bloch sphere for 
a spin squeezed state $|\psi(\kappa)\rangle$,
with $\kappa=N^{1/4}$ corresponding to the squeezing parameter 
$\xi=N^{-1/4}$ 
($N=10$, both the initial state and the state just before detection are 
shown for clarity). 
Thick lines indicate initial uncertainties in $\vec{J}$. 
b) Noise spectra due to L.O. frequency 
fluctuations when free running, when stabilized to unsqueezed 
atoms ($\xi=1$), and when stabilized to spin squeezed atoms 
($\xi=N^{-1/6}$), $N=10^3$ and $\gamma T=10^{-2}$. 
c) Inverse fractional frequency stability $1/\sigma_y$ (arbitrary units) 
vs. Ramsey time for white L.O. noise, $N=10^5$,
with linear  feedback to uncorrelated atoms ($\circ$); linear feedback 
to correlated atoms  ($\diamond$, $\xi=N^{-1/4}$); nonlinear feedback 
to uncorrelated atoms 
($\bullet$); and nonlinear feedback to correlated atoms (filled $\diamond$, 
$\xi=N^{-1/4}$). 
Points: numerical simulations, lines: analytical results.}
\label{fig1}
\end{figure}

When decoherence is negligible, $\langle\delta\phi_O^2\rangle=
\langle\delta\phi_E^2\rangle=0$,
the long term frequency stability is given by 
$\sigma_y(\tau)={\Delta J_y}/{\omega\sqrt{\tau
T}\langle\hat{J}_z\rangle}$ as shown in Refs.  \cite{wineland93,wineland94}.
For an uncorrelated atomic state, the stability improves with increasing 
number of atoms as $N^{-1/2}$ \cite{santarelli99}. 
The maximum possible improvement using spin-squeezed states is a factor of
$N^{-1/2}$, yielding a stability $\sigma_y(\tau)\propto
N^{-1}$ \cite{wineland96}.

The best long term stability is obtained with the
longest possible interrogation time $T$. 
When the interrogation time is limited by
environmental decoherence, the latter cannot be ignored.
This corresponds to the situation considered in Refs.\
\cite{huelga97,kitagawa01}, in which case no substantial improvement is 
possible.
In the practically relevant case where the main source of noise is from
the L.O. \cite{wineland98,wineland00,wineland01_2} the
situation is quite different. In this case it is undesirable to use a very
highly squeezed state with $\Delta J_y\sim 1$ because it has a very large
uncertainty in the $z$-component of the spin $\Delta J_z \sim N$, which
according to Eq.\ (\ref{eq:sigmay}) has a large contribution to the
noise. A moderately squeezed state can, however, lead to a
considerable improvement in the stability. This observation is the main
result of the present Letter.

To find the optimal stability, we first optimize 
(\ref{eq:sigmay}) with respect to the interrogation time. 
Considering uncorrelated atoms first, we have
$\Delta J_y=\sqrt{N}/2$ and $\Delta J_z =0$; Eq. (\ref{eq:sigmay}) then
predicts that $\sigma_y(\tau)$
decreases indefinitely as $1/\sqrt{T}$. To derive Eq. (\ref{eq:sigmay}),
however, we have linearized the expression in Eq. (\ref{eq:nonlinear}),
and this linearization breaks down when the (neglected) cubic term in 
(\ref{eq:linear}) is comparable to the noise term that we retained, i.e., 
when $\delta\phi(t_k)^3\sim\Delta J_y/\langle\hat{J}_z\rangle$. 
In a more careful 
analysis \cite{longpaper} based on Eq. (\ref{eq:linear}), including 
perturbatively the 
nonlinear terms in a stochastic differential equation, we find
the optimal time $\gamma T=(2\Delta
J_y^2/\langle\hat{J}_z\rangle^2)^{1/3}$. At this point the
stability is given by
$\sigma_y(\tau)=\zeta N^{-1/3}\gamma/{\omega\sqrt{\gamma\tau}}$ where
\beq
\zeta=\frac{3}{2^{4/3}}N^{1/3}
\left[
\left(\frac{\Delta J_y}{\langle\hat{J}_z\rangle}\right)^{4/3} 
+\frac{2^{4/3}}{3}
\left(\frac{\Delta J_z}{\langle\hat{J}_z\rangle}\right)^2
\right]^{1/2}.
\label{eq:sqparam}
\eeq 

To evaluate the potential improvement in stability by using squeezed 
states (i.e., the scaling with increasing number of atoms $N$, in the 
limit $N\gg 1$), it is convenient to use a family of states parametrized by 
a small 
number of parameters.  A one-parameter family of states that includes the
uncorrelated state as well as spin squeezed states is given by the
Gaussian states $|\psi(\kappa)\rangle={\mathcal{N}(\kappa)}
\sum_{m}(-1)^{m}e^{-(m/\kappa)^2} |m\rangle$, where $|m\rangle$
are eigenstates of the $J_y$ operator with eigenvalue $m$ and the total 
angular momentum quantum number is $J=N/2$, and
$\mathcal{N}(\kappa)$ is a normalization factor.  The transverse noise for
these states is given by $\Delta J_y=\kappa/2$.
For a large number of atoms
$N\gg 1$, the uncorrelated state 
is well approximated by
$|\psi(\kappa=\sqrt{N})\rangle$, while highly-squeezed states are
obtained when $\kappa\rightarrow 1$.  Within this family of states the
optimal value is $\zeta\simeq 1.42N^{-1/6}$ for $\kappa\simeq 
2^{1/16}N^{1/4}$ 
($\xi\sim N^{-1/4}$)
giving a stability scaling as $N^{-1/2}$. This represents an improvement
by a factor of $N^{1/6}$ compared to uncorrelated states, for which 
$\zeta=3/2^{4/3}$ and the stability scales as $N^{-1/3}$.  
We emphasize that these results are derived assuming a linear feedback 
loop. 

To confirm these predictions, we have made extensive numerical simulations
of the frequency control loop, along the lines of Ref.\ \cite{Audoin98}.
The noise spectrum of the free-running oscillator is defined by 
$S(f)\delta(f+f')=\langle\delta\omega(f)\delta\omega(f')\rangle$, where 
$\delta\omega(f)$ is the Fourier transform of the stochastic process 
$\delta\omega(t)$. 
We generate the corresponding time-series and at the detection
times $t_k=kT$, the accumulated phase $\delta\phi_O(t_k)$ is calculated
and the atomic noise is generated from the probability distributions of
$\mathcal{J}_y$ and $\mathcal{J}_z$. The error signal
$\mathcal{E}(t_k)$ is found and a frequency correction $\Delta\omega(t_k)$
is generated.  
The noise
spectrum of the slaved oscillator, see Fig. 1b, clearly shows that while
for short time scales ($\lesssim T$, high frequencies) the noise is 
given by that of the
free-running oscillator, at longer time scales (lower frequencies) the
oscillator is locked to the atoms and the remaining (white) noise is
determined by the atomic fluctuations. The low-frequency white noise floor 
determines the
long-term stability of the clock and is the quantity we seek to optimize.
In Fig. 1(c) we compare our analytical results with the results of the 
numerical simulations as a function of Ramsey time $T$, and in Fig. 2(a) 
we show the scaling with the number of atoms.
The analytical and numerical approaches are in excellent 
agreement.
\begin{figure}
\epsfig{file=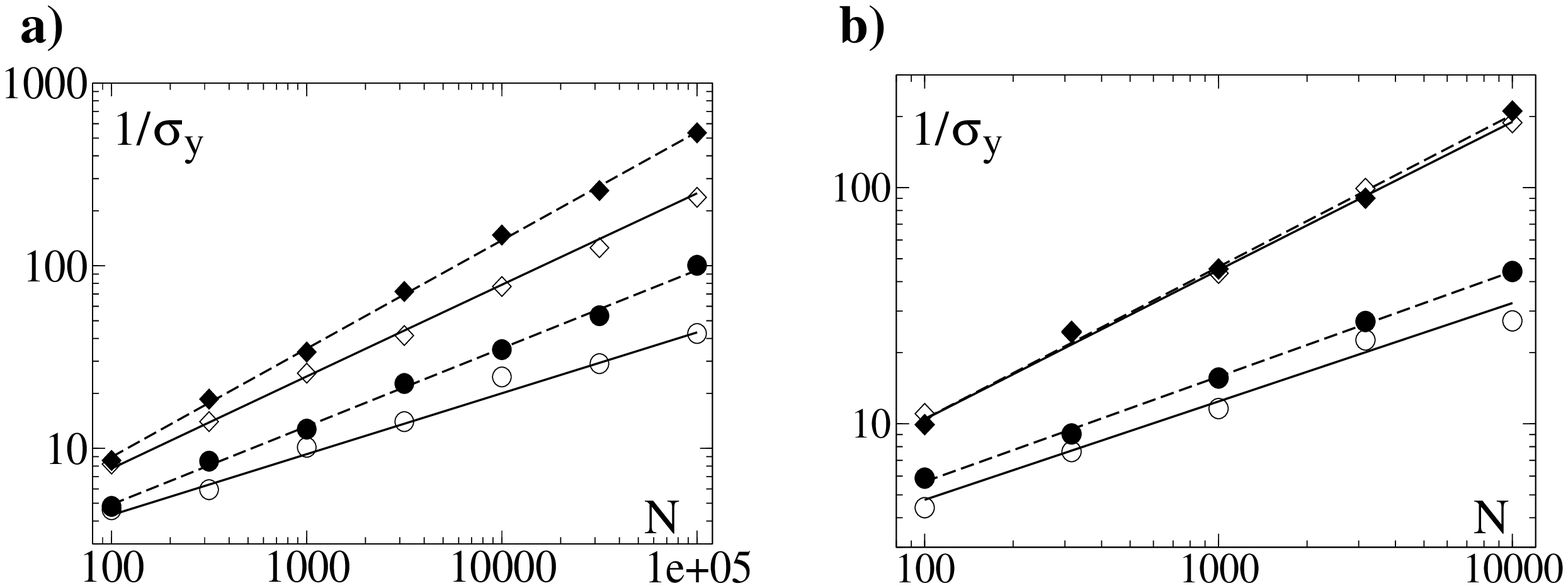,width=8.5cm}
\caption{Inverse fractional frequency stability $1/\sigma_y$ (arbitrary 
units) vs. number of 
atoms $N$, with Ramsey time optimized for a) white noise and b) $1/f$ 
noise.  
Points: numerical simulations, lines: analytical results.
Uncorrelated atoms ($\circ$) and optimal spin 
squeezed atoms ($\diamond$), both for linear feedback (full lines, open 
symbols) and 
nonlinear feedback (dashed lines, filled symbols).}
\label{fig2}
\end{figure}

So far we have assumed linear feedback and white noise;
we now relax these assumptions.
The stability limit identified above is 
mainly determined by the breakdown of the assumption of small (i.e., 
linear) phase fluctuations. In fact, the stability can be improved 
considerably by using a feedback $\Delta\omega$ 
which is a nonlinear function of the error signal $\mathcal{E}$.
To investigate this we have included a nonlinear feedback
$\Delta\omega(t_k)\propto\arcsin[{\mathcal E}(t_k)/J]$
in our numerical simulations.
In Fig. 1(c) it is seen that nonlinear
feedback performs better, and that it extends the validity of Eq.\
(\ref{eq:sigmay}) all the way to $\gamma T\sim 0.1$. 
For larger $\gamma T$, the feedback loop fails, resulting in a rapid 
decrease in stability.
If we optimize the Allan deviation in Eq.\ (\ref{eq:sigmay}) for 
nonlinear feedback, under
the condition $\gamma T\leq 0.1$, we find that the optimally squeezed 
states
have $\Delta J_y\sim N^{1/3}$ ($\xi\sim N^{-1/6}$) resulting in a 
stability scaling as
$N^{-2/3}$. This represents again a relative improvement in scaling of 
$N^{1/6}$
compared to the uncorrelated state for which the stability scales as 
$N^{-1/2}$.
Detailed derivation of these results will be presented elsewhere 
\cite{longpaper}.

The assumption of white noise $\langle\delta\phi^2\rangle=\gamma T$, is 
convenient for theoretical calculations, but in practice 
very-low-frequency noise is likely to have nontrivial spectrum such as 
$1/f$ noise. 
To find the scaling with the number of atoms 
in this situation, we replace $\langle
\delta \phi^2\rangle=\gamma T$ with the behaviour expected for $1/f$ noise:
$\langle \delta \phi^2\rangle\sim (\gamma T)^2$.  Repeating all the
calculations above we again find an improvement by a factor of $N^{1/6}$
by using squeezed states for the nonlinear feedback loop, and a factor of 
$N^{5/24}$ for linear feedback.
In Fig.\ 2b we compare these scaling arguments to the numerical
simulations and the two approaches are seen to be in very good agreement.

To summarize, we have shown that entanglement can provide a significant 
gain in the frequency stability of an atomic clock when it is limited by
the stability of the oscillator used to interrogate the
atoms. The optimal stability  is achieved by using moderately
squeezed states, with a relative improvement that
scales approximately as $N^{1/6}$ with the number of atoms. These
results are in contrast to previous studies using simplified
decoherence models, which found that no practical improvement can be 
achieved with entangled states. 
Finally, we note a few interesting questions raised by our work.
First, it would be interesting to see if there exists special quantum 
states of atoms and feedback mechanisms which optimize the performance 
of the clock.
Second, the present results highlight that it is essential to have
a realistic model of the noise (and possible stabilization mechanism)
present in specific realizations of  quantum information protocols. 
Although the protocol considered in this Letter exploits 
entanglement to stabilize a classical system (the local oscillator), 
it would be interesting to study how similar considerations 
(e.g. $1/f$ noise and collective decoherence) affect protocols such as 
quantum error correction codes \cite{bouwmeester}, which use entanglement 
to stabilize a quantum system and protect it from decoherence.

We are grateful to David Phillips, Ron Walsworth, and David Wineland
for useful discussions and comments on the manuscript. 
This work was supported by the NSF through its CAREER award 
and the grant to ITAMP, by the Packard and Sloan Foundations, 
and by the Danish Natural Science Research Council.



\begin{references}
\bibitem{footnote1}
In practice, $T$ may be limited by various mechanisms, e.g.,
in atomic fountain clocks, $T$ is limited by the available distance the 
atoms may fall under gravity 
\cite{santarelli99}.
\end{references}
\end{document}